\documentclass[journal,9pt]{IEEEtran}

\usepackage{amsmath,amsfonts}
\usepackage{algorithmic}
\usepackage{algorithm}
\usepackage{array}

\usepackage[caption=false,font=normalsize,labelfont=sf,textfont=sf]{subfig}
\usepackage{textcomp}
\usepackage{stfloats}
\usepackage{url}
\usepackage{verbatim}
\usepackage{graphicx}
\usepackage{cite}
\usepackage{hyperref}
\begin{document}

\title{\huge{Multi-frequency Neural Born Iterative Method for Solving 2-D Inverse Scattering Problems}}

\author{\Large{Daoqi Liu, Tao Shan,~\IEEEmembership{Member,~IEEE}, Maokun Li,~\IEEEmembership{Senior Member,~IEEE}, Fan Yang,~\IEEEmembership{Fellow,~IEEE}, \\ and Shenheng Xu,~\IEEEmembership{Member,~IEEE}}
\thanks{Daoqi Liu, Maokun Li, Fan Yang, and Shenheng Xu are with
the Beijing National Research Center for Information Science and Technology (BNRist), Department of Electronic Engineering, Tsinghua University, Beijing 100084, China (e-mail: maokunli@tsinghua.edu.cn).}
\thanks{Tao Shan is with the School of Electronics and Information Engineering, Beihang University, Beijing 100191, China.}}




\maketitle

\begin{abstract}
In this work, we propose a deep learning-based imaging method for addressing the multi-frequency electromagnetic (EM) inverse scattering problem (ISP). By combining deep learning technology with EM physical laws, we have successfully developed a multi-frequency neural Born iterative method (NeuralBIM), guided by the principles of the single-frequency NeuralBIM. This method integrates multitask learning techniques with NeuralBIM's efficient iterative inversion process to construct a robust multi-frequency Born iterative inversion model. During training, the model employs a multitask learning approach guided by homoscedastic uncertainty to adaptively allocate the weights of each frequency's data. Additionally, an unsupervised learning method, constrained by the physical laws of ISP, is used to train the multi-frequency NeuralBIM model, eliminating the need for contrast and total field data. The effectiveness of the multi-frequency NeuralBIM is validated through synthetic and experimental data, demonstrating improvements in accuracy and computational efficiency for solving ISP. Moreover, this method exhibits strong generalization capabilities and noise resistance. The multi-frequency NeuralBIM method explores a novel inversion method for multi-frequency EM data and provides an effective solution for the electromagnetic ISP of multi-frequency data.

\end{abstract}


\section{Introduction}
\IEEEPARstart{T}{he} 
electromagnetic (EM) inverse scattering problem (ISP) is a critical scientific challenge in electromagnetics, involving the inference of unknown scatterers' physical properties within a domain of interest (DOI) from scattered field data \cite{chen2018computational}. This includes determining their shape, size, location, and EM parameters. ISPs are widely used in various fields such as biomedical imaging \cite{abubakar2002BIOimaging} \cite{Fel2020Breast}, nondestructive testing \cite{Salucci2016NDT} \cite{NDTreview}, microwave imaging \cite{shea2010three}, and geophysical exploration \cite{Zhou2023MT,Wang2024SEMI,Lo2010RF}.
However, ISPs are inherently nonlinear and ill-posed \cite{chen2018computational}, meaning that the solution may not exist, may not be unique, or may be highly sensitive to measurement data. These characteristics pose significant challenges to solving ISPs.

Traditional Inverse Scattering Problem (ISP) solutions are primarily categorized into two approaches: qualitative and quantitative methods. Qualitative methods focus on the shape and location of unknown scatterers within the domain of interest (DOI), aiming to produce a generally accurate imaging result without prioritizing the exact values of physical parameters, such as electromagnetic properties. Representative qualitative methods include multiple signal classification (MUSIC) \cite{schmidt1986music}\cite{gruber2004timemusic}, the decomposition of the time-reversal operator (DORT) method \cite{prada1996decomposition}, the linear sampling method (LSM) \cite{colton1996simpleLS}, and the back-propagation (BP) method \cite{belkebir2005superresolution}.
Quantitative methods, on the other hand, strive to obtain precise parameters of the scatterer, including exact values for shape, size, location, and electromagnetic properties. These methods typically involve iterative techniques that formulate the ISP as an optimization problem, constructing a cost function and repeatedly using a forward solver to progressively converge on the solution. Notable quantitative methods include the Born iterative method (BIM) \cite{wang1989iterative}, the distorted Born iterative method (DBIM) \cite{chew1990DBIM}, the variational Born iterative method (VBIM) \cite{zaiping2000variational}, contrast source inversion (CSI) \cite{van1997contrast}, Gauss-Newton inversion method \cite{abubakar2012GN} \cite{Mojabi2009Overview} and the subspace optimization method (SOM) \cite{chen2009subspace}, among others. Compared to qualitative approaches, quantitative methods can derive specific values for electromagnetic properties and other scatterer parameters. However, their effectiveness often depends on the quality of the initial solution, which can limit their performance when dealing with strong scatterers or high-contrast targets.

In recent years, AI technology has developed rapidly, with machine learning, especially deep learning \cite{lecun2015deep}, being successfully applied in many fields such as computer vision (CV) and natural language processing (NLP) \cite{Vou2018deep}\cite{khurana2023natural}. Numerous studies have introduced deep learning technology into EM computation, achieving promising results \cite{AIISP22}. Deep learning-based ISP solving methods leverage the excellent information integration and learning capabilities of deep learning, providing advantages in inversion accuracy and computational efficiency \cite{Shah2017Super,Bayesian,8302596,GAO2022110771,EarlyFDL10332243,wang2021physics,Li2019DeepNIS,Xu2020Deep,Hu2020SDM,Sandhu2021CoSaMP}.
For instance, Zhang et al. proposed a two-step deep learning method involving Frequency Extrapolation and Scatterer Reconstruction \cite{2stepDL}. Pan et al. developed a complex-valued convolutional neural network (CV-CNN) that effectively leverages phase information to address the full-wave nonlinear inverse scattering problem (ISP) \cite{pan2021phase}. Ran et al. applied Convolutional Neural Networks (CNN) to the time-harmonic EM diagnosis of dielectric microstructures \cite{8903073}. Wang et al. proposed GPRI2Net, based on CNN and recurrent neural network (RNN), which can simultaneously reconstruct dielectric constant maps and classify object categories from continuous and long Ground Penetrating Radar (GPR) measurement data \cite{GPRI2Net}. However, many studies, including the aforementioned ones, design deep neural networks for ISP solving in a purely data-driven manner. This approach often results in neural networks being "black boxes", making their internal computations difficult to interpret and the design and training processes challenging \cite{chen2020review}.

Compared to purely data-driven methods, some research has combined data with physical laws, incorporating physical laws into the design of deep neural networks to construct solution methods that integrate physical laws with data-driven approaches \cite{li2023applications}. These methods have yielded excellent results in both efficient EM forward computation \cite{Shan2024GNN}\cite{PHISRL} and effective ISP inversion \cite{GUO2022PDNN}.
In the field of EM forward computation, Shan et al. proposed Physical Information Supervised Residual Learning (PhiSRL) \cite{PHISRL}, a robust and versatile deep learning framework for 2D EM modeling. For ISP solutions, Yao et al. integrated a complex-valued deep convolutional neural network (DConvNet) into the supervised descent method (SDM) to achieve offline training and online "imaging" prediction for EM inverse scattering problems \cite{YaoESDM}. Liu et al. proposed SOM-Net, embedding the Lippmann-Schwinger physical model into neural network structures, continuously updating the induced current and dielectric constant in the sub-network module of SOM-Net \cite{SOMnet}. Hu et al. developed an unsupervised deep learning approach for solving inverse problems, utilizing the computational framework of Physics-Informed neural networks (PINNs) \cite{Wang2023PINN}.

In our previous research, Shan et al. proposed the Neural Born Iterative Method (NeuralBIM) based on the PhiSRL scheme \cite{shan2022neural}. PhiSRL leverages the mathematical connection between the fixed-point iteration method and the residual neural network (ResNet) to solve linear matrix equations by applying CNNs to learn residual updates. Expanding similar design ideas to ISP solving, we developed NeuralBIM, which uses CNNs to learn residual update rules, constructs parameterized functions, and simulates the alternating update process of traditional Born iteration methods while inverting both the total field and scatterers.
However, traditional BIM and NeuralBIM are limited to single-frequency scattering data inversion and are not equipped to handle the inversion of multi-frequency EM data.

Different frequency EM scattering data contain varied information about the scattering object, making multi-frequency data inversion a valuable research topic. Although deep learning has been successfully applied to ISP solving, there is limited work focusing on the inversion of multi-frequency EM data \cite{bao2015inverseMF} \cite{bucci2000ISPMF}. Lin et al. proposed a deep learning-based low-frequency data prediction scheme that integrates information from low-frequency and high-frequency data using frequency-hopping methods to achieve better inversion results \cite{lin2021low}. Li et al. introduced a multi-channel scheme of U-Net CNN to process the BP results of multiple frequencies, obtaining an inversion result that integrates multi-frequency information \cite{CNNMF}. Hu et al. used a residual fully convolutional network (Res-FCN) to learn the scattering data of each frequency point within a wide frequency band, resulting in a model that effectively performs EM inversion of high-contrast scatterers at any frequency within the band \cite{hu2024residual}. Despite these successes, these methods remain purely data-driven, lacking the integration of physical law constraints. There remains substantial room for exploring efficient EM inversion methods that can integrate effective information from multi-frequency scattering data, especially methods that integrate data with physical laws.

This article focuses on the inversion method of multi-frequency EM data and builds upon previous work on NeuralBIM to develop and implement a multi-frequency NeuralBIM approach \cite{shan2022neural}. This method integrates multi-task learning technology with NeuralBIM's efficient iterative inversion technique, resulting in a multi-frequency NeuralBIM iterative inversion model. A multi-task learning method guided by homoscedastic uncertainty is employed to adaptively allocate the weight of each frequency data during the iterative computation in training. We use a physics-guided unsupervised learning method to train the multi-frequency NeuralBIM model, which is constrained by the physical laws of the inverse scattering problem, eliminating the need for contrast and total field data. The final multi-frequency NeuralBIM method effectively extracts key information from multi-frequency EM scattering data and achieves robust inversion results for complex scattering objects with high contrast, demonstrating strong generalization and noise immunity.

The structure of this article is as follows: Section \ref{section2} introduces the theoretical basis of ISP, and Section \ref{section3} explains the principles and architecture of single-frequency and multi-frequency NeuralBIM. Section \ref{section4} details the multi-task learning method and unsupervised learning mode used. Section \ref{section5} provides numerical results, demonstrating the effectiveness of the proposed multi-frequency NeuralBIM in solving EM inverse scattering problems. Finally, Section \ref{section6} summarizes the article and outlines the potential impact and contributions of the proposed method in the field of EM inversion.

\section{Inverse Scattering Problems}\label{section2}
\begin{figure}[htbp]
  \centering
  \includegraphics[width=2.5in]{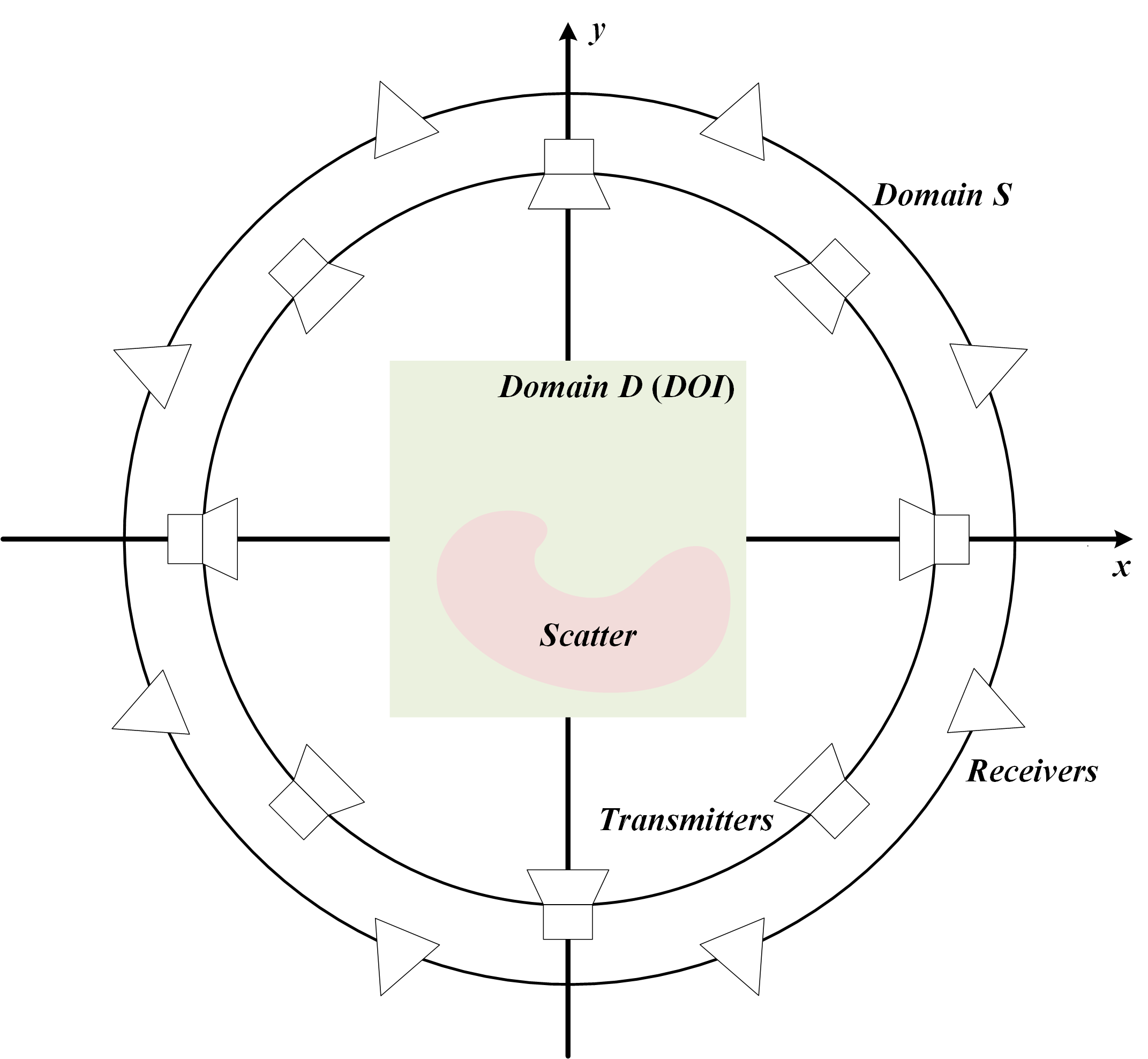}
  \caption{ Modeling of inverse scattering problems in 2-D cases. }
  \label{fig1}
\end{figure}

\hyperref[fig1]{Fig. 1} illustrates the inverse scattering problem (ISP) in a two-dimensional region, where the unknown scatterers are located within the DOI, while the background consists of free space. The objective of the ISP is to reconstruct the scatterer contrast parameter, denoted as $\chi$. The source and receiving antenna for the transverse magnetic(TM) wave are distributed around the domain, and the incident field $E_{i}$ and total field $E_{t}$ at position $\bf{r}$ satisfy the following equations \cite{chen2018computational}:
\begin{equation}\label{eq1}
  E_{t}(\mathbf{r})= E_{i}(\mathbf{r})+ k_b^2 \int_{D}G_D(\mathbf{r},\mathbf{r'})\chi(\mathbf{r'})E_{t}(\mathbf{r'})d\mathbf{r'}, \mathbf{r}\in D
\end{equation}

where DOI is represented as domain $D$. The scattered field $E_{s}$ received by the antenna at position $\mathbf{r_s}$ in the receiving area $S$ is given by \cite{chen2018computational}:
\begin{equation}\label{eq2}
  E_{s}(\mathbf{r_s})= \int_{D}G_S(\mathbf{r_s},\mathbf{r'})\chi(\mathbf{r'})E_{t}(\mathbf{r'})d\mathbf{r'}, \mathbf{r_s}\in S
\end{equation}
where $k_b$ is the wave number, $G_S$ and $G_D$ are the free-space Green's functions of domain S and domain D, respectively. Using the method of moments (MoM), the DOI is subdivided into discrete elements, allowing for the discretization of the calculation process of (\ref{eq1}) and (\ref{eq2}). (\ref{eq1}) and (\ref{eq2}) can be expressed in matrix form:
\begin{equation}\label{eq3}
  \mathbf{(I-G_D\chi)E_{t}=E_{i}}
\end{equation}
\begin{equation}\label{eq4}
  \mathbf{E_{s}=G_S\chi E_{t}}.
\end{equation}

\section{Neural Born Iterative Method}\label{section3}
\subsection{Traditional Born Iterative Method}
Traditional Born Iterative Method (TBIM) is a common method for solving ISPs. TBIM first solves the optimization problem:
\begin{equation}\label{eq5}
  \min_{\mathbf{\chi}} || \mathbf{E_{s}-G_S\chi E_{t}} ||^2
\end{equation}
to obtain an updated value of $\chi$. The updated value $\chi_{update}$ is obtained and then used to find the updated value of $E_{t}$
\begin{equation}\label{eq6}
  \mathbf{E_{t}^{update}=(I-G_D\chi_{update})^{-1}E_{i}}
\end{equation}
Repeat the above iterative update process until the iteration result meets the requirements.

\subsection{Single-Frequency Neural Born Iterative Method}

NeuralBIM enhances the traditional BIM by incorporating the design concept of Physical Information Supervised Residual Learning (PhiSRL) \cite{PHISRL}. This method integrates CNN with the algorithm, combining a ResNet \cite{he2016RESNET} and physical laws to learn the rules for iterative updates in BIM. A physically-guided deep neural network simulates the calculation process of the TBIM, resulting in improved convergence and iterative updates of $\chi$ and $E_{t}$.

The process of updating $\chi$ in NeuralBIM is as follows:
\begin{equation}\label{eq7}
  \begin{aligned}
  \mathbf{Res_{E_{s}}^k}&= \mathbf{E_{s}-G_S\chi^k E_{t}^k} \\
  \mathbf{\chi^{k+1}}&=\mathbf{\chi^k}+\mathcal{N}_{\chi}^{k+1}(\mathbf{Res_{E_{s}}^k\oplus \chi^k}, \mathbf{\Theta^{k+1}_{\chi}})
  \end{aligned}
\end{equation}
where $\mathbf{Res_{E_{s}}^k}$ denotes the residual component of the scattered field $E_{s}$. Concurrently, $\mathcal{N}_{\chi}^{k+1}$ signifies the CNN associated with $\chi$, as employed within the framework of the model. $\mathbf{\Theta^{k+1}_{\chi}}$ refers to the internal parameter set specific to the $\chi-CNN$. The $\mathbf{Res_{E_{s}}^k}$ and $\mathcal{N}_{\chi}^{k+1}$ are utilized as input parameters for the $\chi-CNN$, leading to the iterative update of $\chi$. $\oplus$ signifies tensor concatenation. Similarly, the procedure for updating $E_{t}$ using NeuralBIM is as follows:
\begin{equation}\label{eq8}
  \begin{aligned}
    \mathbf{Res_{E_{i}}^k}&= \mathbf{E_{i}-(I-G_D\chi^{k+1}) E_{t}^k} \\
    \mathbf{E_{t}^{k+1}}&=\mathbf{E_{t}^k}+\mathcal{N}_{E_{t}}^{k+1}(\mathbf{Res_{E_{i}}^k}, \mathbf{\Theta^{k+1}_{E_{t}}})
    \end{aligned}
\end{equation}
where $\mathbf{Res_{E_{i}}^k}$ denotes the residual component of the incident EM field $E_{i}$, while $\mathcal{N}_{E_{t}}^{k+1}$ represents the $E_{t}-CNN$ employed in the model. Moreover, $\mathbf{\Theta^{k+1}_{E_{t}}}$ encapsulates the distinct set of internal parameters that are unique to the $E_{t}-CNN$. The NeuralBIM algorithm systematically executes an iterative computational procedure, as delineated in the respective equations, persisting in this loop until the pre-established criteria for convergence are satisfactorily met. The procedural architecture of the NeuralBIM algorithm is delineated in Algorithm \ref{alg1}.

\begin{algorithm}[H]
  \caption{Neural Born Iterative Method}\label{alg:alg1}
  \begin{algorithmic}
  \STATE 
  \STATE \textbf{Initialization:} $\mathbf{E_{t}^{0}}=\mathbf{E_{i}}$, $\mathbf{\chi^{0}=0}$, $k=0$, $k_{max}$
  \STATE \textbf{while} $k<k_{max}$ \textbf{do}
  \STATE \hspace{0.5cm} \textbf{step 1:} $\mathbf{Res_{E_{s}}^k}= \mathbf{E_{s}-G_S\chi^k E_{t}^k}$
  \STATE \hspace{0.5cm} \textbf{step 2:} $\mathbf{\chi^{k+1}}=\mathbf{\chi^k}+\mathcal{N}_{\chi}^{k+1}(\mathbf{Res_{E_{s}}^k\oplus \chi^k}, \mathbf{\Theta^{k+1}_{\chi}})  $
  \STATE \hspace{0.5cm} \textbf{step 3:} $\mathbf{Res_{E_{i}}^k}= \mathbf{E_{i}-(I-G_D\chi^{k+1}) E_{t}^k}  $
  \STATE \hspace{0.5cm} \textbf{step 4:} $\mathbf{E_{t}^{k+1}} =\mathbf{E_{t}^k}+\mathcal{N}_{E_{t}}^{k+1}(\mathbf{Res_{E_{i}}^k}, \mathbf{\Theta^{k+1}_{E_{t}}}) $
  \STATE \hspace{0.5cm} \textbf{step 5:} $k=k+1 $
  \STATE \textbf{end while}
  \STATE \textbf{Output:} $\mathbf{\chi^{k_{max}}}$,$\mathbf{E_{t}^{k_{max}}}$
  \end{algorithmic}
  \label{alg1}
  \end{algorithm}

\subsection{Multi-Frequency Neural Born Iterative Method}
According to the principle and architecture of NeuralBIM, we improve it to achieve multi-frequency data inversion. For multi-frequency data, the incident field, scattered field, and total field data of the $i$th frequency are represented as $\mathbf{E_{i}(f_i)}$, $\mathbf{E_{s}(f_i)}$ and $\mathbf{E_{t}(f_i)}$.  According to (\ref{eq7}) and (\ref{eq8}), we can calculate the residual of the scattered field $\mathbf{Res_{E_{s}}(f_i)}$ and the residual of the total field $\mathbf{Res_{E_{t}}(f_i)}$ at the $i$th frequency.  For cases with $n$ frequency data, similar to the single-frequency NeuralBIM, the process of updating $\chi$ in multi-frequency NeuralBIM is as follows:
\begin{equation}
  \mathbf{Res_{E_{s}}^k(f_i)}= \mathbf{E_{s}(f_i)-G_S^i\chi^k E_{t}^k(f_i)} 
\end{equation}
\begin{equation}\label{eq9}
  \mathbf{Res_{E_{s}}^k}=\mathbf{Res_{E_{s}}^k(f_1)} \oplus \cdots \oplus\mathbf{Res_{E_{s}}^k(f_n)}
\end{equation}
\begin{equation}
  \mathbf{\chi^{k+1}}=\mathbf{\chi^k}+\mathcal{N}_{\chi}^{k+1}(\mathbf{Res_{E_{s}}^k} \oplus \chi^k, \mathbf{\Theta^{k+1}_{\chi}}),
\end{equation}
where $\mathbf{Res_{E_{s}}^k}$ is the concatenation of the $\mathbf{E_{s}}$ residuals at each frequency, and $\oplus$ is the concatenation of two tensors. $G_S^i$ denotes Green's function in free space for the $i$th frequency. $\mathcal{N}_{\chi}^{k+1}$ represents the $\chi-CNN$ for multi-frequency data, and $\mathbf{\Theta^{k+1}_{\chi}}$ is the internal parameter set of the network. Inputting the residual of multi-frequency data and $\chi$ results in an iterative update of $\chi$. 

Similarly, the process of updating $\mathbf{E_{t}}$ for multi-frequency NeuralBIM is as follows:
\begin{equation}\label{eq10}
  \mathbf{Res_{E_{i}}^k(f_i)}= \mathbf{E_{i}(f_i)-(I-G_D^i\chi^{k+1}) E_{t}^k(fi)} 
\end{equation}
\begin{equation}
  \mathbf{Res_{E_{i}}^k}=\mathbf{Res_{E_{i}}^k(f_1)} \oplus \cdots \oplus\mathbf{Res_{E_{i}}^k(f_n)}
\end{equation}
\begin{equation}
  \mathbf{E_{t}^{k+1}}=\mathbf{E_{t}^k}+\mathcal{N}_{E_{t}}^{k+1}(\mathbf{Res_{E_{i}}^k}, \mathbf{\Theta^{k+1}_{E_{t}}}),
\end{equation}
where $\mathbf{Res_{E_{i}}^k}$ is the concatenation of the $\mathbf{E_{i}}$ residuals at each frequency, and $\oplus$ is the concatenation of two tensors. $G_D^i$ denotes Green's function in free space for the $i$th frequency. $\mathcal{N}_{E_{t}}^{k+1}$ represents the $E_{t}-CNN$ for multi-frequency data, and $\mathbf{\Theta^{k+1}_{E_{t}}}$ is the internal parameter set of the network.
The procedural architecture of the NeuralBIM algorithm is delineated in Algorithm \ref{alg2}.

\begin{algorithm}[H]
  \caption{Multi-Frequency Neural Born Iterative Method}\label{alg:alg2}
  \begin{algorithmic}
  \STATE 
  \STATE \textbf{Initialization:} $\mathbf{E_{t}^{0}(f_i)}=\mathbf{E_{i}(f_i)}$, $\mathbf{\chi^{0}=0}$, $k=0$, $k_{max}$
  \STATE \textbf{while} $k<k_{max}$ \textbf{do}
  \STATE \hspace{0.5cm} \textbf{step 1:} $\mathbf{Res_{E_{s}}^k(f_i)}= \mathbf{E_{s}(f_i)-G_S^i\chi^k E_{t}^k(f_i)}$
  \STATE \hspace{0.5cm}  $\mathbf{Res_{E_{s}}^k} =\mathbf{Res_{E_{s}}^k(f_1)} \oplus \cdots \oplus\mathbf{Res_{E_{s}}^k(f_n)}$
  \STATE \hspace{0.5cm} \textbf{step 2:} $\mathbf{\chi^{k+1}}=\mathbf{\chi^k}+\mathcal{N}_{\chi}^{k+1}(\mathbf{Res_{E_{s}}^k\oplus \chi^k}, \mathbf{\Theta^{k+1}_{\chi}})  $
  
  \STATE \hspace{0.5cm} \textbf{step 3:} $\mathbf{Res_{E_{i}}^k(f_i)}= \mathbf{E_{i}(f_i)-(I-G_D^i\chi^{k+1}) E_{t}^k(f_i)}  $
  \STATE \hspace{0.5cm}  $\mathbf{Res_{E_{i}}^k}=\mathbf{Res_{E_{i}}^k(f_1)} \oplus \cdots \oplus\mathbf{Res_{E_{i}}^k(f_n)}$
  \STATE \hspace{0.5cm} \textbf{step 4:} $\mathbf{E_{t}^{k+1}} =\mathbf{E_{t}^k}+\mathcal{N}_{E_{t}}^{k+1}(\mathbf{Res_{E_{i}}^k}, \mathbf{\Theta^{k+1}_{E_{t}}}) $
  \STATE \hspace{0.5cm} \textbf{step 5:} $k=k+1 $
  \STATE \textbf{end while}
  \STATE \textbf{Output:} $\mathbf{\chi^{k_{max}}}$,$\mathbf{E_{t}^{k_{max}}}$
  \end{algorithmic}
  \label{alg2}
  \end{algorithm}

\begin{figure*}[htbp]
  \centerline{\includegraphics[width=2.0\columnwidth]{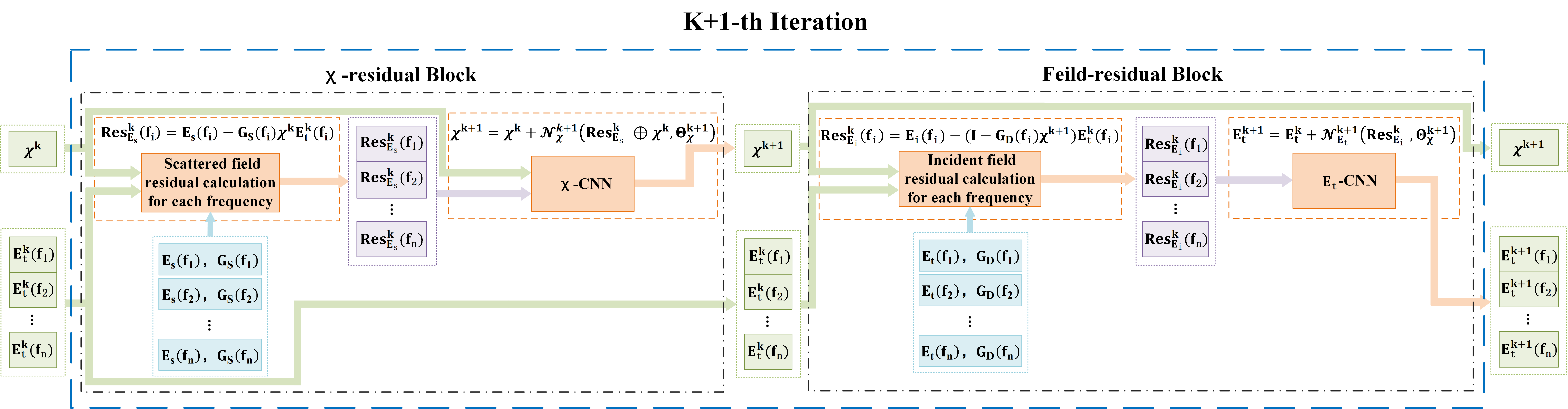}}
  \caption{The calculation process for the \(k+1\)-th iteration. Calculate the scattered field residual $\mathbf{Res_{E_{s}}^k(f_i)}$ of each frequency based on the contrast $\mathbf{\chi^k}$ and the total field $\mathbf{E_{t}^k}$ from the \( k \)-th iteration. Subsequently, use $\chi-CNN$ to update the contrast $\mathbf{\chi^{k+1}}$. Next, compute the residual $\mathbf{Res_{E_{i}}^k(f_i)}$ of the incident field at each frequency based on $\mathbf{\chi^{k+1}}$ and $\mathbf{E_{t}^k}$, and update the total field $\mathbf{E_{t}^{k+1}}$ using $E_{t}-CNN$ to obtain the results for the \(k+1\)-th iteration.}
  \label{fig2}
\end{figure*}

\hyperref[fig2]{Fig. 2} illustrates the k+1-th iteration process of the multi-frequency NeuralBIM. During each iteration, the data for $\mathbf{\chi}$ and $\mathbf{E_{t}}$ are input into the iteration update module, where they are subsequently updated through computation. As delineated in (\ref{eq9}) to (\ref{eq14}), in the k+1-th iteration, the value of $\mathbf{\chi^k}$ is initially updated via the $\chi$-residual Block iteration to obtain $\mathbf{\chi^{k+1}}$. Subsequently, $\mathbf{\chi^{k+1}}$ and $\mathbf{E_{t}^k}$ are input into the Field-residual block to update the value of $\mathbf{E_{t}^{k+1}}$. This completes one iteration cycle. The multi-frequency NeuralBIM employed in this study comprises seven iterative update modules cascaded, as depicted in \hyperref[fig2]{Fig. 2}. The initial data values are refined through seven iterations of the NeuralBIM to yield the final result.

\begin{figure}[htbp]
  \centering
  \includegraphics[width=0.95\linewidth]{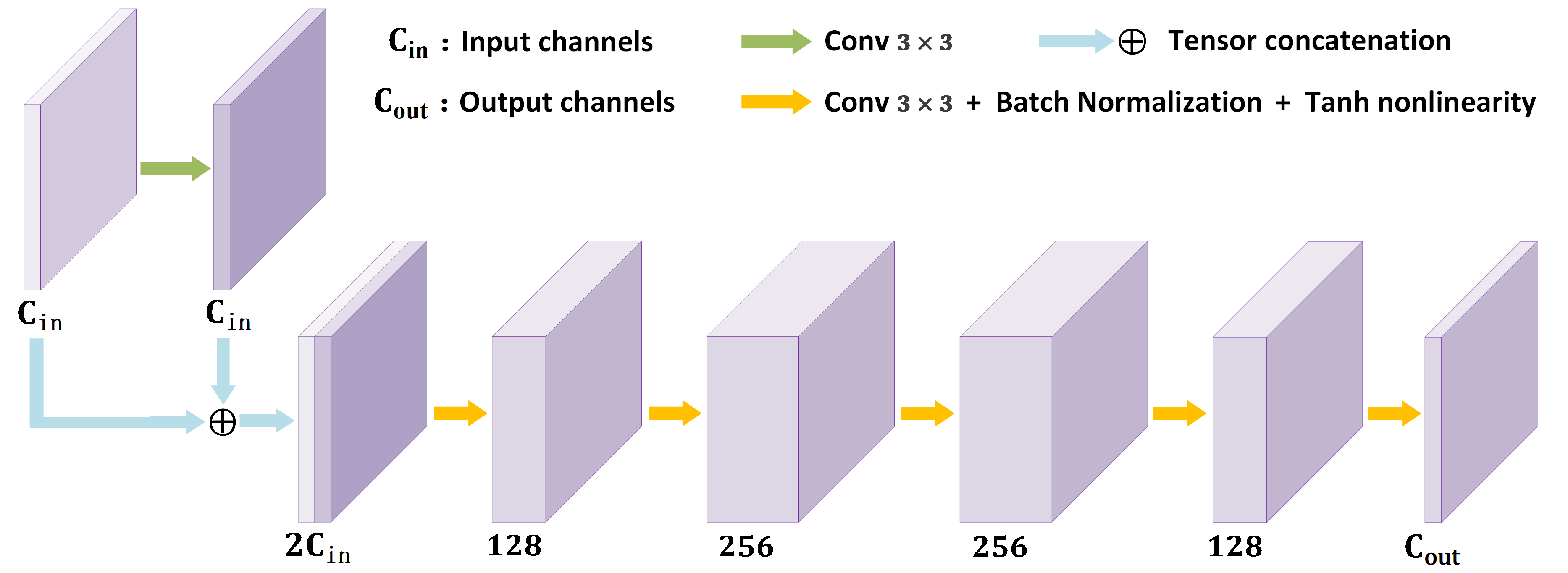}
  \caption{ Specific architectures of $\chi-CNN$ and $E_{t}-CNN$. The architectures of $\chi-CNN$ and $E_{t}-CNN$ are the same but their respective parameter sets are different.}
  \label{fig3}
\end{figure}

In the iteration update module of neuralBIM, $\chi-CNN$ and $E_{t}-CNN$ are employed to compute the iterative update values for $\chi$ and $E_{t}$, respectively. Both networks share an identical compositional architecture, as depicted in \hyperref[fig3]{Fig. 3}, comprising $3\times3$ convolutional layers, batch normalization (BN) layers, and hyperbolic tangent (Tanh) nonlinearity layers. The number of input channels is denoted by $C_{in}$, while the number of output channels is represented by $C_{out}$. In the context of a multi-frequency NeuralBIM with $n$ frequencies, the input channel dimension for $\chi-CNN$ is given by $C_{in}=2n+2$, and the output channel dimension is $C_{out}=2$. For $E_{t}-CNN$, both the input and output channel dimensions are equal, with $C_{in}=C_{out}=2$.

\section{Multitask Learning for Multi-Frequency Data}\label{section4}
Multitask learning has found widespread application across various domains of deep learning \cite{zhang2018overview}, including CV \cite{zhang2013robust} \cite{kendall2018multi}, NLP \cite{ICASSP2015}, and bioinformatics \cite{MTLforBIO}. It addresses the challenge of optimizing a model with respect to multiple objectives simultaneously \cite{SurveyMTL}. In this study, we employ multitask learning for the inversion computation of multi-frequency data, leading to the development of a multi-frequency NeuralBIM model. This model integrates the information conveyed by each frequency data, thereby enhancing its predictive capability. In the context of the multi-frequency NeuralBIM model, the iterative computation for each frequency data constitutes a distinct task. All these tasks share the initial input data $\mathbf{\chi^0}$, as well as the frequency-specific input data $\mathbf{E_{i}(f_i)}$, $\mathbf{E_{s}(f_i)}$ and \( \mathbf{E_{t}^0(f_i)} \). Moreover, they collectively produce a common output \(\mathbf{\chi^{out}}\) and a frequency-specific output \( \mathbf{E_{t}^{out}(f_i)} \).

\subsection{Objective Function of Multitask Learning}

The loss associated with the $i$th task, herein referred to as the inverse calculation loss of the $i$th frequency, is denoted by \(Loss_i\). Consequently, the objective function is articulated as the weighted aggregation of the task-specific losses:

\begin{equation}\label{eq11}
  Obj = \sum_i \omega_i Loss_i 
\end{equation}

Addressing the equilibrium among task-specific losses emerges as a pivotal concern. The distribution of weights, \(\omega_i\), corresponding to each loss exerts a considerable influence on both the efficacy of the model and the computational outcomes. The manual specification of each task's loss is fraught with challenges in achieving optimal results, as each iteration and adjustment of weights demands substantial temporal and resource investments. Hence, the exploration of methodologies for ascertaining optimal weight distribution is warranted.

A multi-task learning paradigm that leverages homoscedastic uncertainty, embracing probabilistic modeling principles, is proposed for the derivation of optimal task weights \cite{kendall2018multi}. This approach employs homoscedastic uncertainty as a criterion for the apportionment of loss weights across distinct tasks, facilitating ongoing adjustment and learning of these weights throughout the training phase. The foundation of this method is the maximization of the Gaussian likelihood estimate predicated on homoscedastic uncertainty, culminating in the formulation of a multi-task loss function. Specifically, within the context of a neural network model tasked with \(n\) distinct functions, denoted as
\begin{equation}\label{eq12}
  D_{out}^i = N(D_{in}, \mathbf{\theta}),
\end{equation}
where \(D_{in}\) represents the input data, \(\mathbf{\theta}\) encapsulates the network parameters, and \(D_{out}^i\) signifies the output of the $i$th task, for \(i = 1,2, \cdots, n\).

The likelihood is conceptualized as a Gaussian distribution, its mean reflective of the model's output

\begin{equation}\label{eq14}
  p(D_{out}^i, D_{in}, \mathbf{\theta}) = \frac{1}{2\pi\sigma_i^2} \exp\left(-\frac{\|D_{truth}^i -D_{out}^i\|^2}{2\sigma_i^2}\right).
\end{equation}
\begin{equation}\label{eq15}
  p(D_{out}^1,\cdots,D_{out}^n, D_{in}, \mathbf{\theta}) = \prod_i p(D_{out}^i, D_{in}, \mathbf{\theta}). 
\end{equation}
wherein \(\sigma_i\) symbolizes the covariance indicative of the network model's uncertainty pertaining to the $i$th task and $D_{truth}^i$ represents the ground truth. To transform (\ref{eq14}) into its maximum likelihood form, the log-likelihood of the model can be maximized by:

\begin{equation}\label{eq17}
  \begin{aligned}
    &\log p(D_{out}^1,\cdots,D_{out}^n, D_{in}, \mathbf{\theta})\\ &\propto \sum_i -\frac{1}{2\sigma_i^2} \| D_{truth}^i -D_{out}^i \|^2 - \log\sigma_i . 
  \end{aligned}
\end{equation}

Based on (\ref{eq17}) , using homoscedasticity uncertainty to weight the losses of different tasks, the objective function for multi-task learning can be formulated as:
\begin{equation}\label{eq18}
  Obj = \sum_i \frac{1}{2\sigma_i^2}Loss_i + \log\sigma_i . 
\end{equation}
Here, homoscedasticity uncertainty \(\sigma_i\) represents a learnable loss weight parameter that adaptively adjusts the loss weight for each task during the training process. In the context of the multi-frequency NeuralBIM model, the output result for each frequency data computes the corresponding loss $Loss_i$. According to (\ref{eq18}), homoscedasticity uncertainty \(\sigma_i\) is introduced to adjust the loss weight of each frequency during training.

\subsection{Unsupervised Learning Method}
The multi-frequency NeuralBIM employs an unsupervised learning method for model training. Unsupervised learning confers benefits such as obviating the need for labeled data, enhancing flexibility and versatility, and diminishing preprocessing requisites. Within the unsupervised learning method, (\ref{eq3}) and (\ref{eq4}) pertaining to the ISP are utilized to govern the training of the multi-frequency NeuralBIM model. During the training process, the true values of the total field \(\mathbf{E_{t}}\), and the contrast \(\mathbf{\chi}\), remain undisclosed. It is postulated that the Green's function, the scattered field \(\mathbf{E_{s}}\), and the incident field \(\mathbf{E_{i}}\), are known entities in the ISP. The training model comprehensively assimilates existing physical laws and information, enabling the multi-frequency NeuralBIM to concurrently satisfy (\ref{eq3}) and (\ref{eq4}) across various frequencies. Specifically, for the \(i\)th frequency, the loss function \(Loss_i\) is defined as:

\begin{equation}\label{eq19}
  Loss_i = R_{E_{i}}^i + R_{E_{s}}^i
\end{equation}

where \(R_{E_{i}}^i\) is related to the residual of the incident field obtained based on (\ref{eq3}):
\begin{equation}\label{eq20}
  R_{E_{i}}^i= \frac{1}{N_{E_{i}}(f_i)}\left\|(\mathbf{I} - \mathbf{G_D^i \chi^{out}}) \mathbf{E_{t}^{out}(f_i)} - \mathbf{E_{i}(f_i)}\right\|^2_F
\end{equation}

Analogously, \(R_{E_{s}}^i\) can be procured based on (\ref{eq4}), and total variation (TV) regularization is incorporated:

\begin{equation}\label{eq21}
  R_{E_{s}}^i = \frac{1}{N_{E_{s}}(f_i)}\left\|(\mathbf{E_{s}(f_i)} - \mathbf{G_S^i \chi^{out} E_{t}^{out}})\right\|^2_F + \alpha \left\|\nabla \mathbf{\chi^{out}}\right\|_1
\end{equation}

where \(N_{E_{i}}(f_i)\) and \(N_{E_{s}}(f_i)\) denote the number of elements in \(\mathbf{E_{i}(f_i)}\) and \(\mathbf{E_{s}(f_i)}\) respectively. $\mathbf{\chi^{out}}$ and $\mathbf{E_{t}^{out}}$ are the inverse results of $\mathbf{\chi}$ and $\mathbf{E_{t}}$ obtained by iterative computation, respectively. The TV regularization term, \(\|\nabla \mathbf{\chi^{out}}\|_1\), introduced in (\ref{eq20}), serves to stabilize the training process and ensure commendable boundary delineation and homogeneity of the inversely reconstructed scatterer. \(\alpha\) is set to $0.00001$, \(\|\cdot\|_F\) signifies the Frobenius norm, and \(\|\cdot\|_1\) represents the L1 norm.

Finally, by amalgamating (\ref{eq18}) and (\ref{eq19}), the objective function for the multi-frequency neuralBIM, designed to integrate multi-frequency data information through multi-task learning in an unsupervised learning framework, can be derived:

\begin{equation}\label{eq22}
  Obj = \sum_i \frac{1}{2\sigma_i^2}\left(R_{E_{i}}^i + R_{E_{s}}^i\right) + \log\sigma_i , 
\end{equation}
where \(R_{E_{i}}^i\) and \(R_{E_{s}}^i\) are defined in (\ref{eq20}) and (\ref{eq21}) respectively. $\sigma_i$ denotes a learnable parameter representing the uncertainty associated with the result of $i$th frequency.
This objective function enables the multi-frequency neuralBIM to conduct multi-task learning across various frequencies in an unsupervised learning method.

\section{Numerical Results}\label{section5}

In this section, we first construct multi-frequency NeuralBIM with three frequencies: 3GHz, 4GHz, and 5GHz. Then we use synthetic and experimental data to verify the inversion performance of multi-frequency NeuralBIM. We use the Pytorch library to construct multi-frequency NeuralBIM and the Adam optimizer \cite{kingma2014adam} to optimize the neural network parameters. The computation of multi-frequency NeuralBIM is performed using three Nvidia V100 GPUs.

\subsection{Synthetic Data Inversion}

\begin{figure}[htbp]
  \centering
  \includegraphics[width=0.95\linewidth]{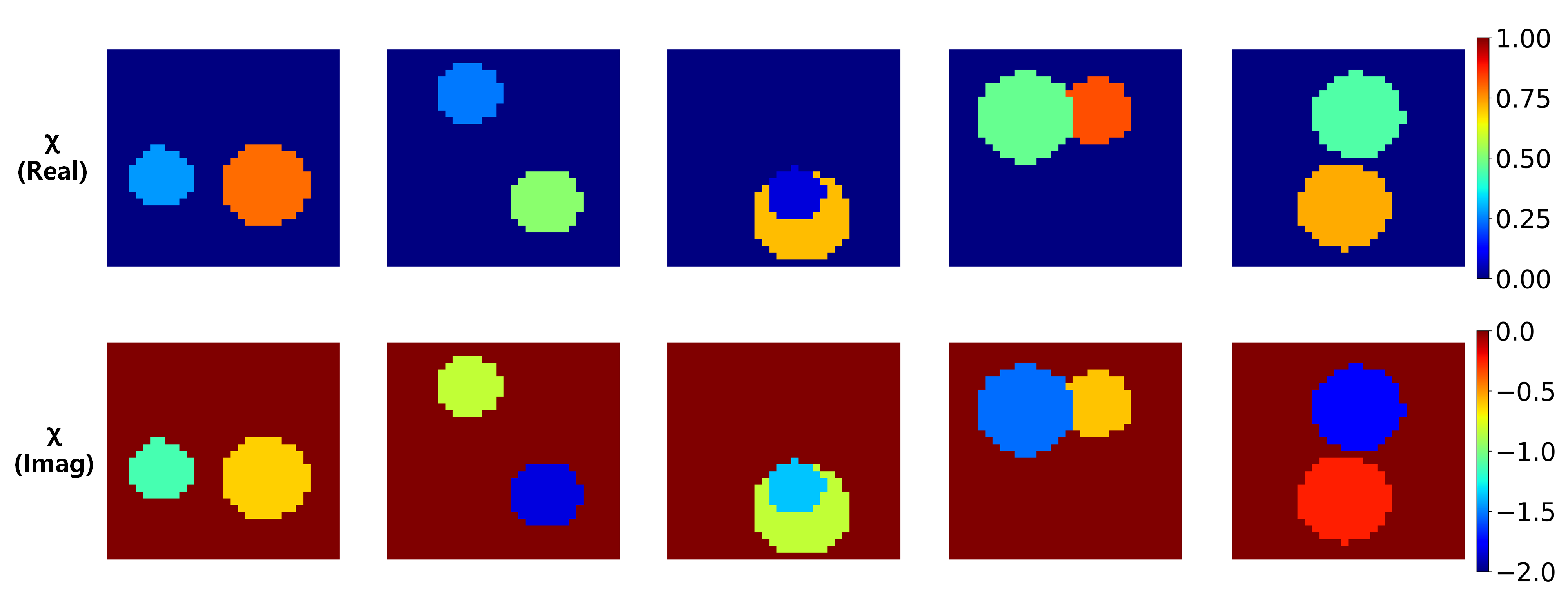}
  \caption{Examples of synthetic data scatterers, with the size and contrast of cylindrical composites randomly distributed within the range shown in Table \ref{table1}. }
  \label{fig4}
\end{figure}

As shown in Figure 1, the domain $D$ (DOI) in the ISP is set to $0.15m\times0.15m$ and divided into a $32\times32$ grid. The $32$ transmitters and receivers are evenly distributed on a circle with a radius of $1.57m$ centered on the DOI center, forming domain $S$. The scatterers generating data are two uniform cylindrical combinations randomly distributed in the DOI, with their sizes and contrasts randomly taken from the ranges shown in Table \ref{table1}.  The background in the DOI is free space. A sample of some syntactic data scatterers is shown in \hyperref[fig4]{Fig. 4}. After randomly generating scatterers, the EM scattering data of the scatterers under incident fields at 3GHz, 4GHz, and 5GHz are calculated using MoM, namely $\mathbf{E_{i}}$ and $\mathbf{E_{s}}$ in (\ref{eq3}) and (\ref{eq4}). One scatterer and its EM scattering data at three frequencies constitute a data sample. A total of 10,000 data samples were generated, with $80\%$ randomly selected to form the training set and the remaining $20\%$ forming the testing set. The initial value of the scatterer contrast $\mathbf{\chi^0}$is obtained using the BP method, and the initial value of the total field $\mathbf{E_{t}^0}$ is considered to be the incident field $\mathbf{E_{i}}$.
\begin{table}[htbp]
  \caption{Scatterer Parameters of Synthetic Data Inversion\label{table1}}
  \centering
  \begin{tabular}{|c|c|c|c|}
  \hline
  Cylinder & Radius /$m$ & Contrast(Real) & Contrast(Imag)\\
  \hline
  $A$ & $\left[0.015,0.035\right]$ & $\left[0,1\right]$ & $\left[-1,0\right]$\\ 
  \hline
  $B$ & $\left[0.015,0.035\right]$ & $\left[0,1\right]$ & $\left[-2,-1\right]$\\ 
  \hline
  \end{tabular}
\end{table}

\begin{figure}[htbp]
  \centering
  \includegraphics[width=0.95\linewidth]{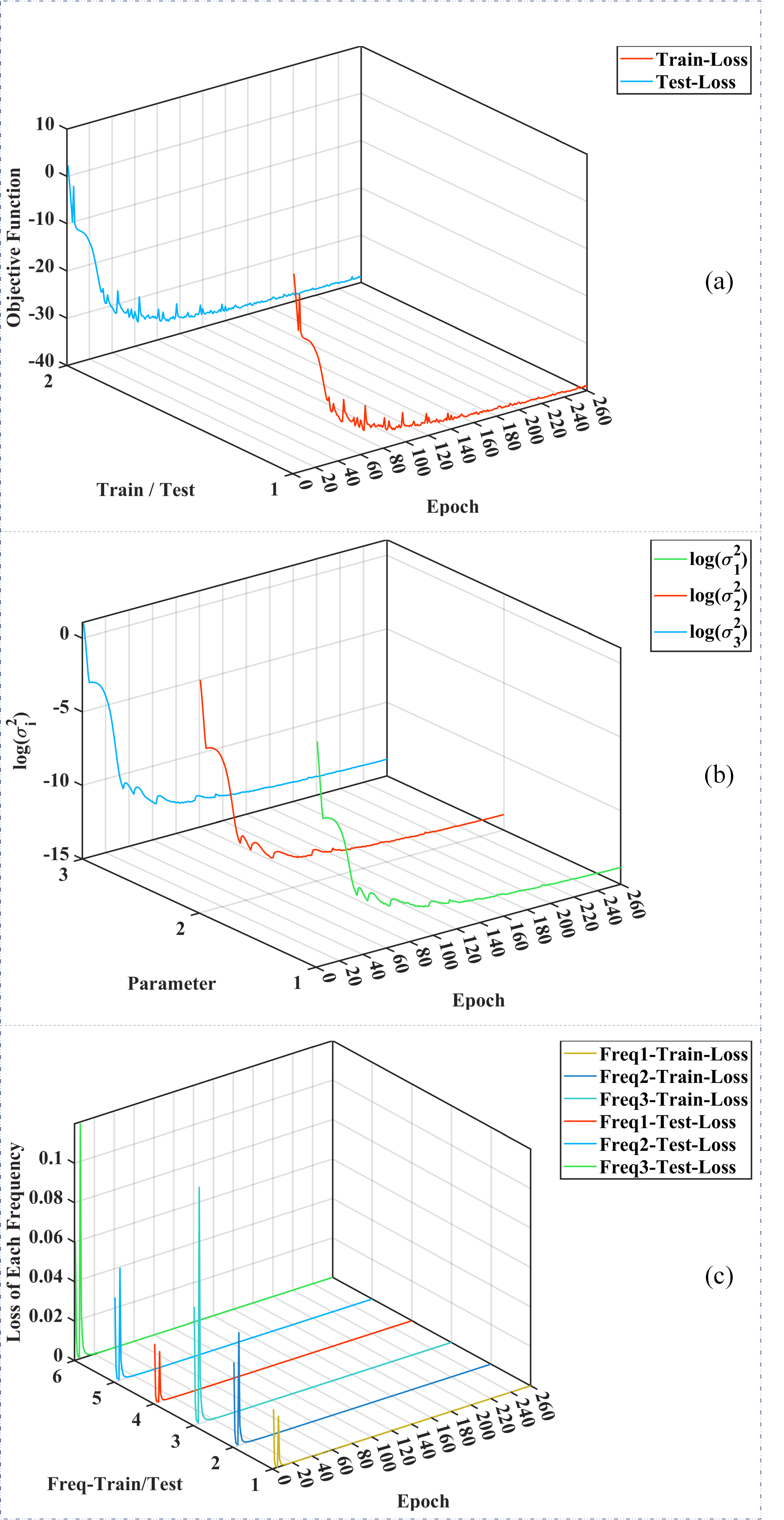}
  \caption{Convergence curve of multi-frequency NeuralBIM. (a) Variation curve of the objective function. (b) Variation curve of the weight parameter $\sigma_i$, where the ordinate represents $log(\sigma_i^2)$. (c) Loss values at three frequencies. Freq1, Freq2 and Freq3 represent 3GHz, 4GHz and 5GHz, respectively. }
  \label{fig5}
\end{figure}

The model underwent training for a total of 260 epochs. \hyperref[fig5]{Fig. 5(a)} illustrates the variation curve of the objective function as defined in (\ref{eq22}), demonstrating that the training process of the model ultimately converges, with the loss values for both the training and testing sets being essentially identical. \hyperref[fig5]{Fig. 5(b)} presents the loss values at various frequencies during the training process, calculated using (\ref{eq19}). The loss values and their variations at three distinct frequencies during the training are observed to be largely consistent. \hyperref[fig5]{Fig. 5(c)} depicts the variation in the learnable parameters $\sigma_i$ (as introduced in (\ref{eq22})), which are employed to modulate the loss weights of different tasks throughout the training process. It is evident that the weight parameter $\sigma_i$, derived based on the homoscedasticity uncertainty, can adaptively adjust the loss weights of each task during training, thereby enabling the model to fully utilize multi-frequency information. The diverse information presented in \hyperref[fig5]{Fig. 5} signifies the stability of multi-frequency neuralBIM training and the dependability of the model architecture.

\begin{figure}[htbp]
  \centering
  \includegraphics[width=0.95\linewidth]{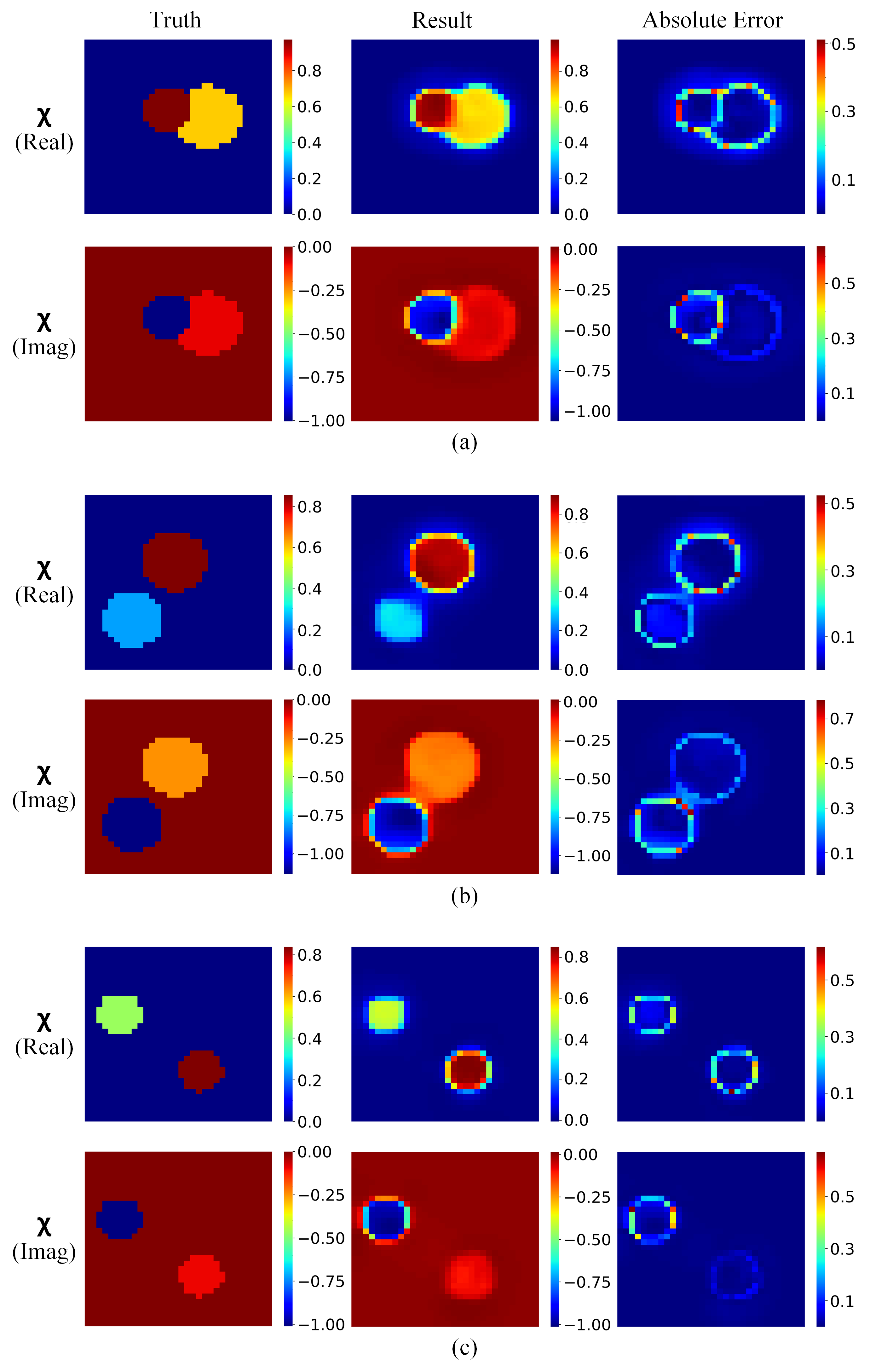}
  \caption{Comparison between the ground truth contrast values and the multi-frequency neuralBIM inversion results selected randomly from the testing set. (a) Scatterers composed of two overlapping cylinders. (b) Scatterers composed of two adjacent cylinders. (c) Scatterers composed of two separated cylinders. Each row from left to right represents the ground truth value, inversion result, and mean absolute error, respectively. }
  \label{fig6}
\end{figure}

A random selection of results from the testing set was used to plot the comparison between the ground truth contrast values and the model inversion results, along with the mean absolute error(MAE) between the two, as depicted in \hyperref[fig6]{Fig. 6}. The inversion results closely resemble the ground truth values in terms of shape and contrast, with errors primarily concentrated around the boundary of the scatterers, indicating satisfactory overall inversion performance.

\begin{figure}[htbp]
  \centering
  \includegraphics[width=0.98\linewidth]{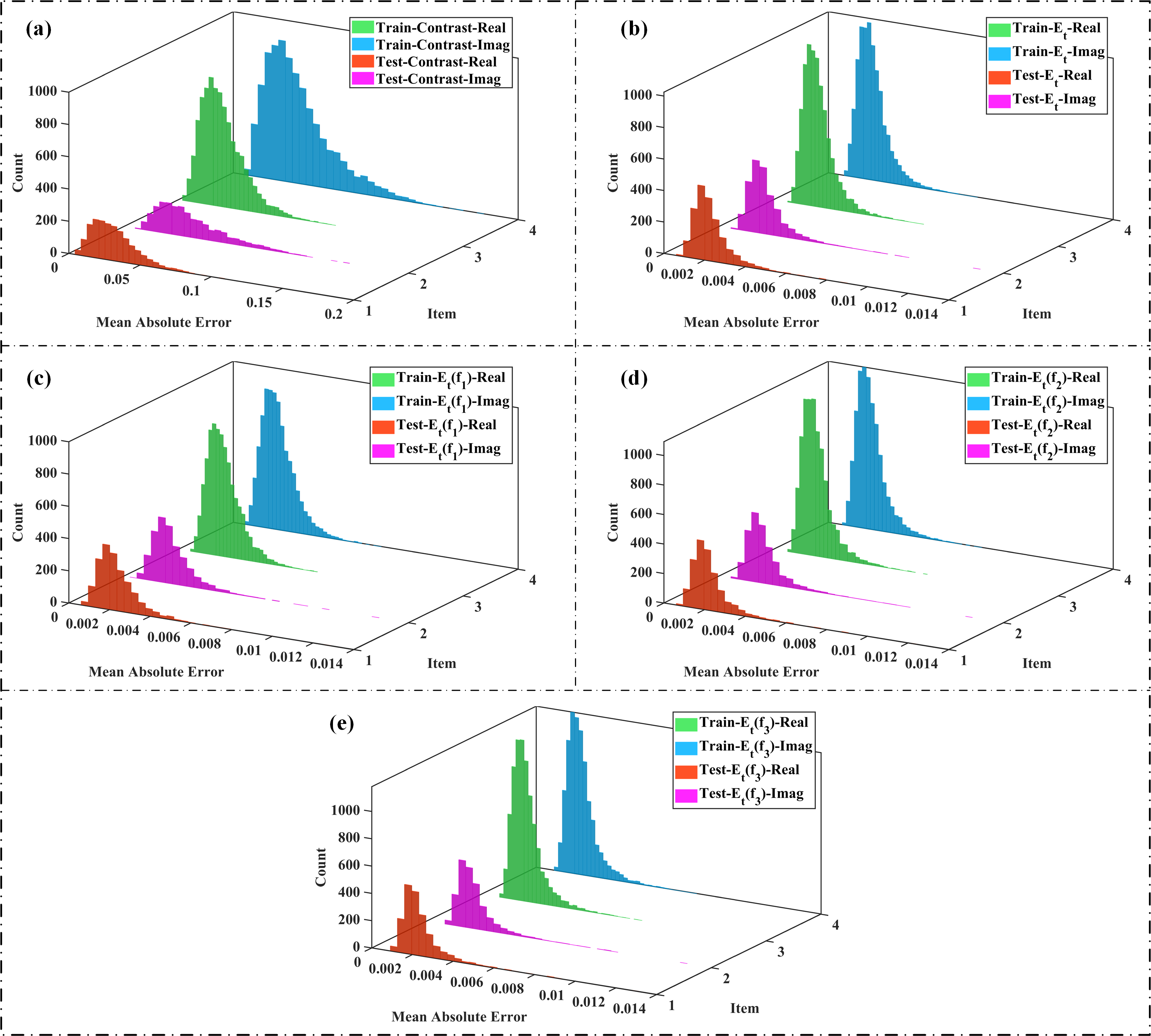}
  \caption{ MAE histogram for training and testing data samples. (a) MAE histogram of contrast. (b) MAE histogram of $\mathbf{E_{t}}$. (c),(d) and (e) MAE histogram of $\mathbf{E_{t}}$ at three frequencies. $f_1$, $f_2$ and $f_3$ represent 3GHz, 4GHz and 5GHz, respectively.}
  \label{fig7}
\end{figure}

A histogram representing the MAE distribution between the training and testing datasets is statistically illustrated in \hyperref[fig7]{Fig. 7}. \hyperref[fig7]{Fig. 7(a) and (b)} display the MAE distribution of the real and imaginary parts of the scatterer contrast $\mathbf{\chi}$ and the total field $\mathbf{E_{t}}$ calculation results, respectively. \hyperref[fig7]{Fig. 7(c), (d)  and (e)} show the MAE distribution of $\mathbf{E_{t}}$ calculation at three different frequencies. Table \ref{table2} provides the mean and standard deviation of the aforementioned MAE results. The mean and standard deviation of the MAE between $\mathbf{\chi}$ and $\mathbf{E_{t}}$ of the testing and training sets are found to be essentially identical, as are the mean and standard deviation of the MAE of each frequency $\mathbf{E_{t}}$ between the testing and training sets. This consistency aligns with the variations in the objective function and loss at each frequency depicted in \hyperref[fig5]{Fig. 5}, indicating an absence of overfitting in the model training. The information on MAE distribution reflects the effective performance of the inversion calculation and the stability of the model training.

\begin{table}[htbp]
  \caption{Mean and Standard Deviation(STD) of MAE in Multi-frequency NeuralBIM Inversion Results\label{table2}}
  \centering
  \begin{tabular}{|c|c|c|}
  \hline
  Item & MAE-Mean/STD(Real) & MAE-Mean/STD(Imag)\\
  \hline
  Contrast-Train$^{(1)}$ & $0.0306$ / $0.0147$ & $0.0454$ / $0.0227$ \\ 
  \hline
  Contrast-Test$^{(2)}$ & $0.0312$ / $0.0154$ & $0.0461$ / $0.0228$ \\
  \hline
  $\mathbf{E_{t}}$-Train$^{(3)}$ & $0.0022$ / $7.4\times10^{-4}$ & $0.0022$ / $7.4\times10^{-4}$ \\ 
  \hline
  $\mathbf{E_{t}}$-Test$^{(4)}$ &$0.0023$ / $9.4\times10^{-4}$ & $0.0023$ / $9.4\times10^{-4}$\\
  \hline
  $\mathbf{E_{t}(f_1)}$-Train & $0.0022$ / $8.1\times10^{-4}$ & $0.0022$ / $8.1\times10^{-4}$\\
  \hline
  $\mathbf{E_{t}(f_1)}$-Test  &$0.0023$ / $9.9\times10^{-4}$ & $0.0023$ / $9.9\times10^{-4}$\\   
  \hline
  $\mathbf{E_{t}(f_2)}$-Train & $0.0021$ / $7.3\times10^{-4}$ & $0.0021$ / $7.3\times10^{-4}$\\
  \hline
  $\mathbf{E_{t}(f_2)}$-Test  &$0.0022$ / $9.5\times10^{-4}$ & $0.0022$ / $9.5\times10^{-4}$\\   
  \hline
  $\mathbf{E_{t}(f_3)}$-Train & $0.0022$ / $7.3\times10^{-4}$ & $0.0022$ / $7.2\times10^{-4}$\\
  \hline
  $\mathbf{E_{t}(f_3)}$-Test  &$0.0023$ / $9.4\times10^{-4}$ & $0.0023$ / $9.4\times10^{-4}$\\   
  \hline
  \end{tabular}
  (Real): Real part\quad
  (Imag): Imaginary part \\
  $^{(1)}$ Contrast of training set\quad
  $^{(2)}$ Contrast of testing set\\
  $^{(3)}$ Total field of training set\quad
  $^{(4)}$ Total field of testing set\\
  ${\mathbf{E_{t}(f_1)}}$: 3GHz total field \quad
  ${\mathbf{E_{t}(f_2)}}$: 4GHz total field \\
  ${\mathbf{E_{t}(f_3)}}$: 5GHz total field
\end{table}

\begin{figure}[htbp]
  \centering
  \includegraphics[width=0.98\linewidth]{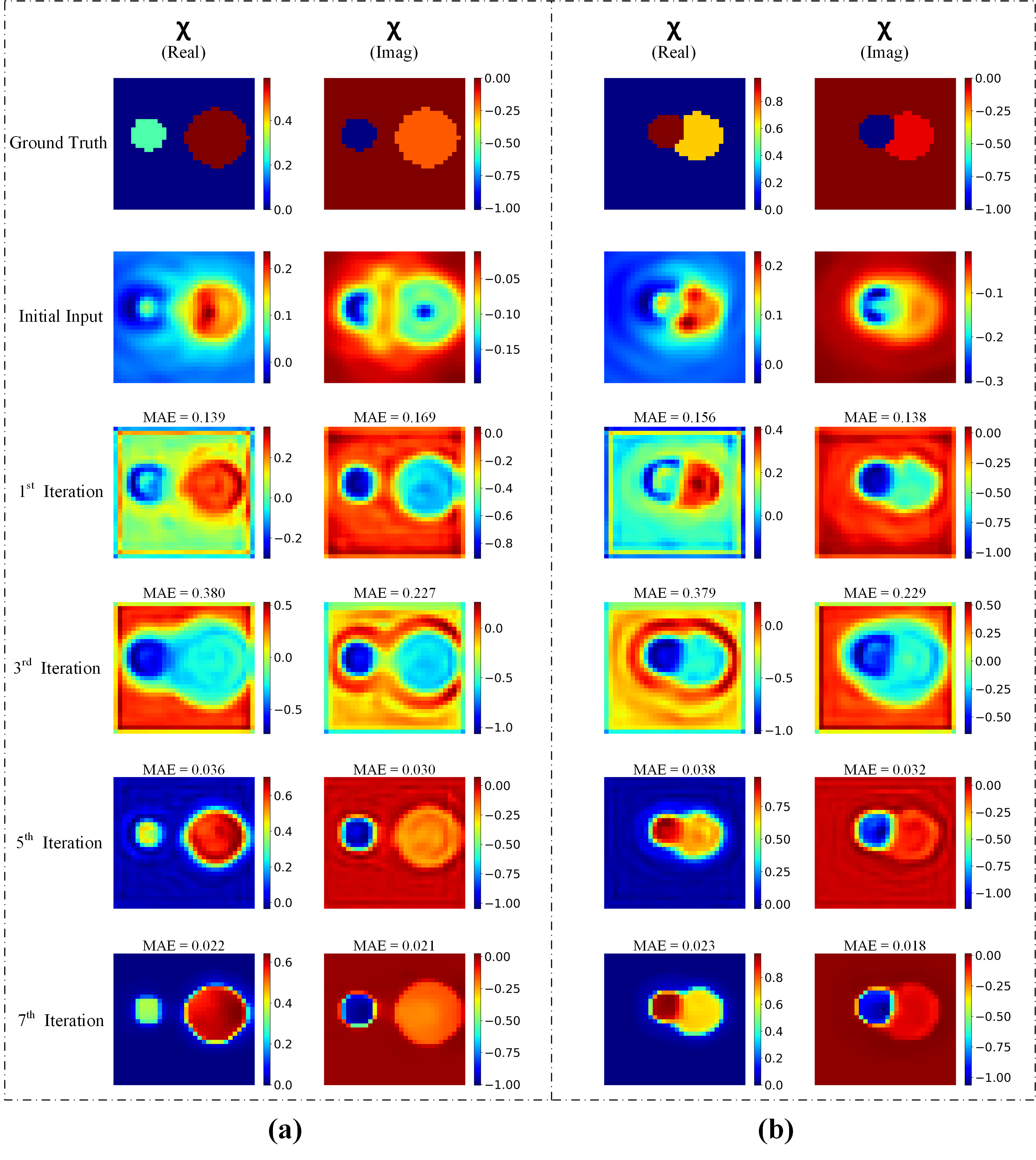}
  \caption{ The iterative calculation process of scatter contrast. The ground truth value of the scatterer contrast, the initial input, and the inversion results after the $1^{st}$, $3^{rd}$, $5^{th}$, and $7^{th}$ iterations of the model are plotted. (a) Scatterers composed of two separated cylinders. (b) Scatterers composed of two overlapping cylinders. }
  \label{fig8}
\end{figure}

\hyperref[fig8]{Fig. 8} illustrates the iterative calculation process of multi-frequency NeuralBIM model inversion and marks contrast MAE in the intermediate results for various iterations. The initial value of the scatterer contrast, obtained using the BP method, serves as the model input. It is observed that as the number of model iterations increases, the inversion results progressively converge towards the ground truth values. The iterative variation process of the inversion results for different scatterers exhibits a similar pattern.

\begin{figure}[htbp]
  \centering
  \includegraphics[width=0.95\linewidth]{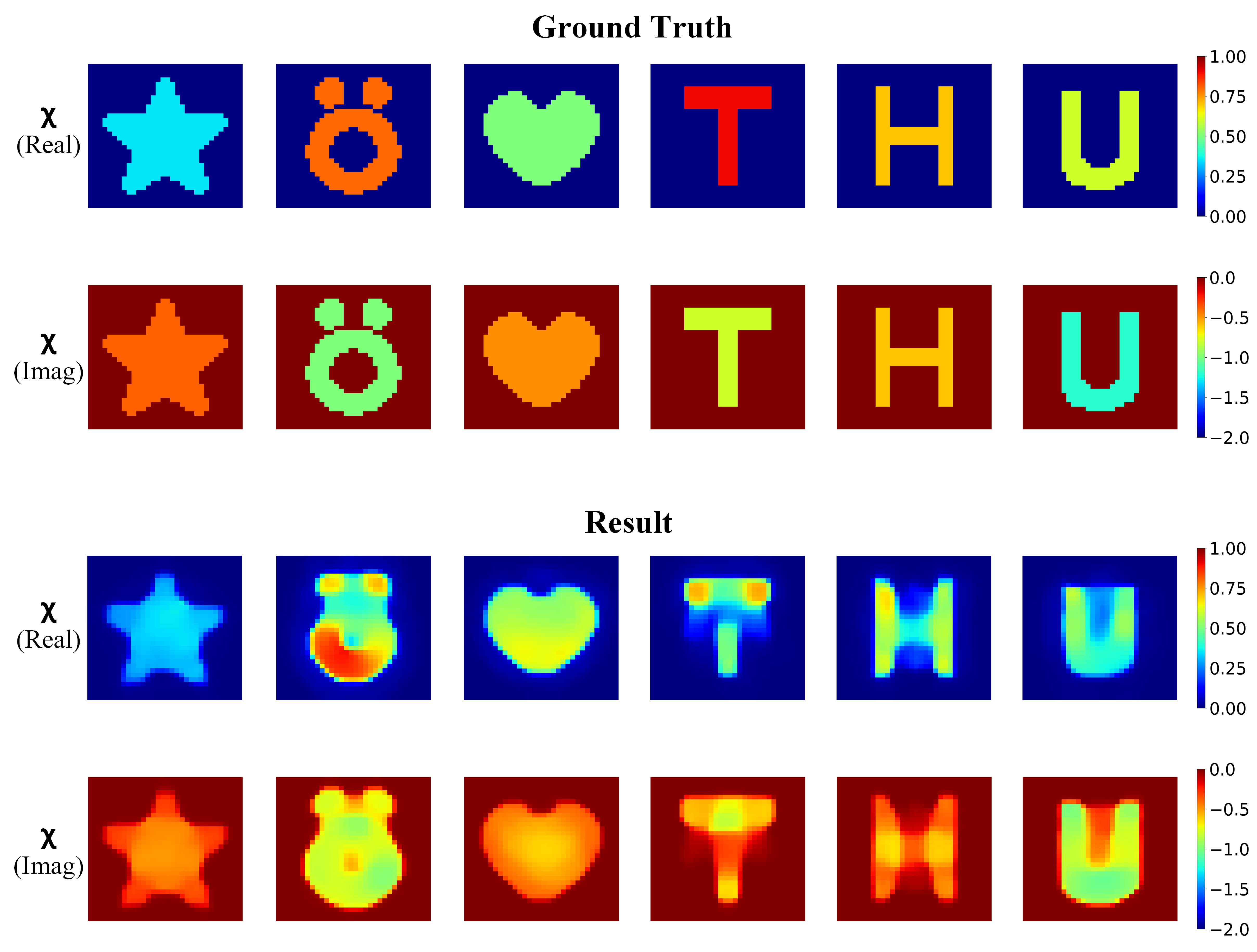}
  \caption{ Verify the generalization ability of multi-frequency NeuralBIM model using scatterers of various shapes not available in the training set. }
  \label{fig9}
\end{figure}

To assess the generalization capability of the model, a set of uniquely shaped scatterers, not present in the training set, was selected to evaluate the inversion performance of the multi-frequency NeuralBIM model on this set of scatterers. \hyperref[fig9]{Fig. 9} displays the shapes and contrast values of these six scatterers, alongside the model's inversion results for them. The inversion results closely match the ground truth in terms of shape and contrast, with errors primarily concentrated on small detailed structures, indicating that the multi-frequency NeuralBIM model maintains a robust inversion performance for scatterers of various shapes not encountered in the training set.

\begin{figure}[htbp]
  \centering
  \includegraphics[width=0.95\linewidth]{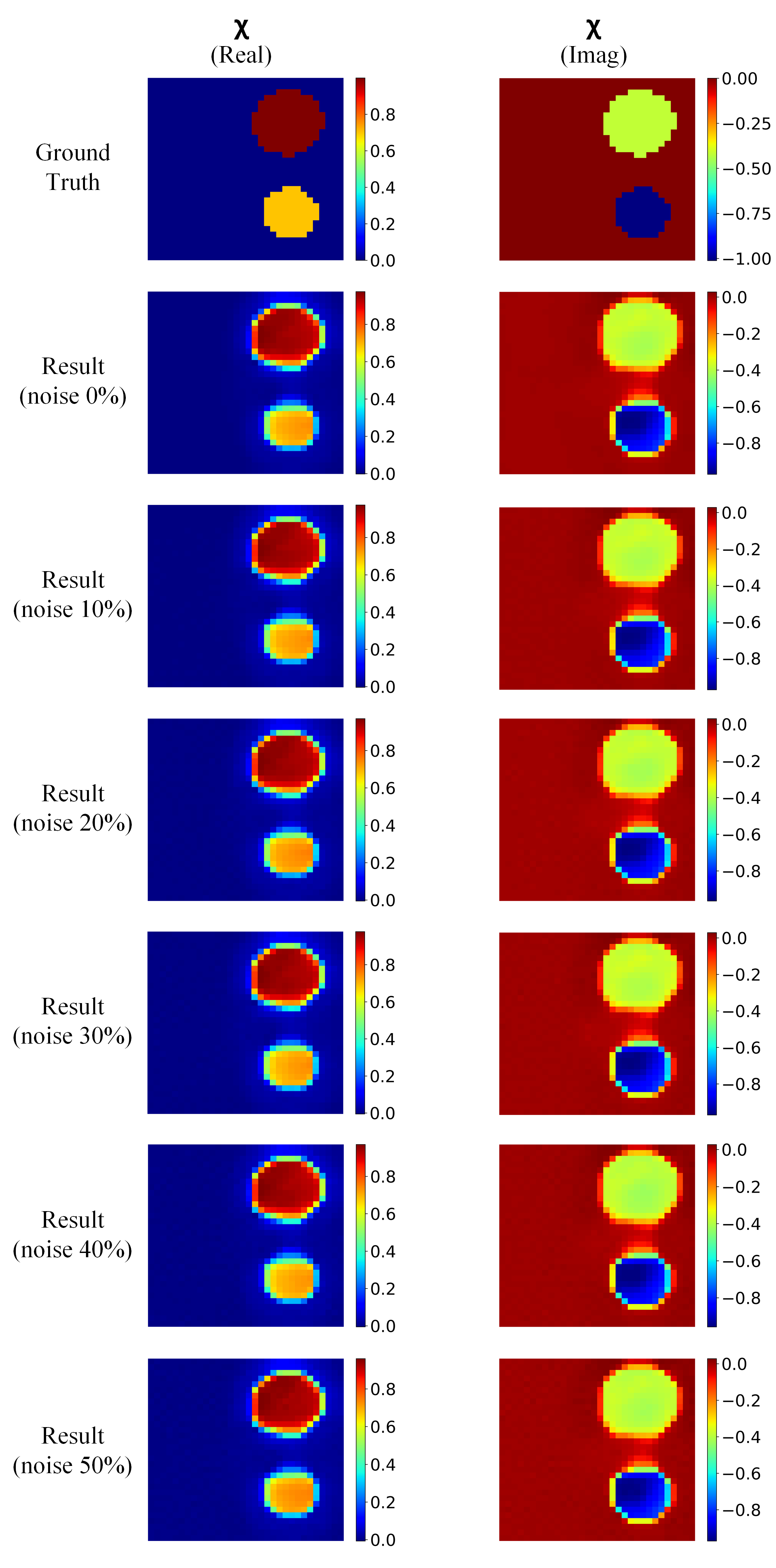}
  \caption{ Ground truth and inversion results of multi-frequency NeuralBIM under different noise levels. Ground truth and the results of noise levels from $0\%$ to $50\%$ are drawn from top to bottom. }
  \label{fig10}
\end{figure}

\begin{figure}[htbp]
  \centering
  \includegraphics[width=0.95\linewidth]{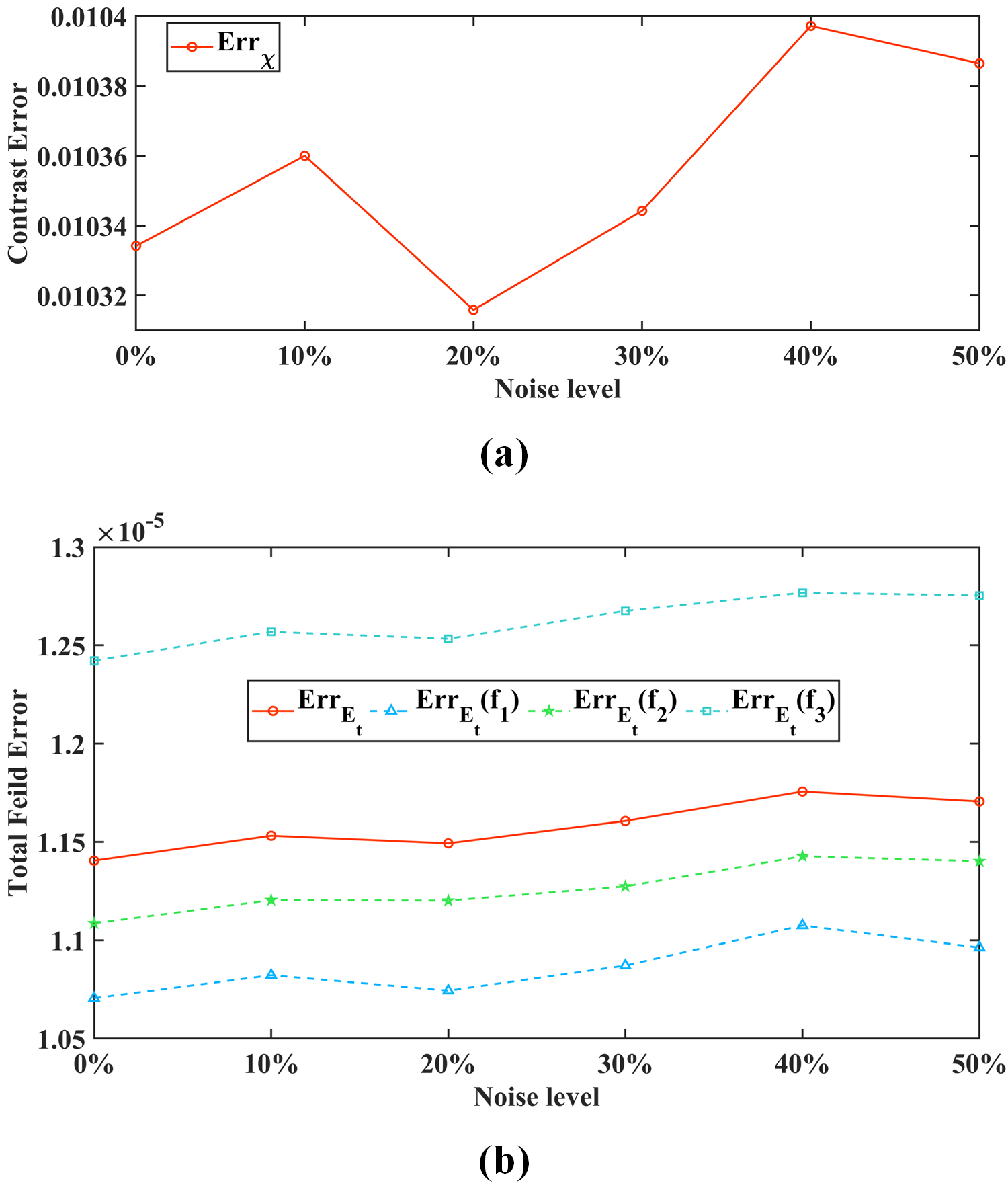}
  \caption{ Model inversion errors for Contrast and total field under different noise levels. (a) Model inversion error for Contrast. (b) Model inversion error for total field, including the average total error for three frequencies (red line) and the total field error for each frequency. $f_1$, $f_2$ and $f_3$ represent 3GHz, 4GHz, and 5GHz, respectively. }
  \label{fig11}
\end{figure}

To evaluate the model's noise resilience, 40 samples were randomly selected from the test set, and varying levels of Gaussian white noise were applied to the samples to observe the alterations in the inversion results. \hyperref[fig10]{Fig. 10} presents the input of $\mathbf{E_{s}}$ at different noise levels, as well as the corresponding inversion results. It is evident that the multi-frequency NeuralBIM model still yields satisfactory inversion results even at a $50\%$ noise level, and the model output remains stable under various noise levels. The following equation is defined to calculate the error of the output results for contrast and $\mathbf{E_{t}}$:

\begin{equation}\label{eq23}
  \begin{aligned}
    Err_{\chi} &= {\frac{1}{N_{\chi}}}\left\|\mathbf{\chi^{truth}} - \mathbf{\chi^{out}}\right\|^2_F \\
    Err_{E_{t}} &= {\frac{1}{N_{E_{t}}}}\left\|\mathbf{E_{t}^{truth}} - \mathbf{E_{t}^{out}}\right\|^2_F \\
  \end{aligned}
\end{equation}

where \(\mathbf{\chi^{truth}}\) and \(\mathbf{E_{t}^{truth}}\) are the ground truth of contrast and $\mathbf{E_{t}}$, respectively, and \(\mathbf{\chi^{out}}\) and \(\mathbf{E_{t}^{out}}\) are the model outputs of contrast and $\mathbf{E_{t}}$, respectively. \(N_{\chi}\) and \(N_{E_{t}}\) are the number of elements in \(\chi\) and \(\mathbf{E_{t}}\), respectively. For model outputs with different noise levels, the inversion errors of contrast and \(\mathbf{E_{t}}\) were calculated using (\ref{eq23}), as shown in \hyperref[fig11]{Fig. 11}. The multi-frequency NeuralBIM demonstrates stable and robust output for inputs with varying noise levels, exhibiting strong noise resistance.

\subsection{Experimental Data Inversion}
The performance of the multi-frequency NeuralBIM was validated using experimental data obtained from Institut Fresnel \cite{geffrin2005free}. Specifically, the data pertained to the TM polarization case for the \textit{FoamDielInt} and \textit{FoamDielExt} models. The scatterers comprised a small cylinder with $\epsilon_r=3.3$ and a diameter of $0.31m$, and a large cylinder with $\epsilon_r=1.45$ and a diameter of $0.8m$. The frequencies employed by the multi-frequency NeuralBIM were 3 GHz, 4 GHz, and 5 GHz. The DOI setting and grid division in the ISP were consistent with those used for synthetic data inversion. The experimental measurement configuration consisted of 8 transmitters and 241 receivers, arranged on a circle with a radius of 1.57 m centered on the DOI. Given the disparity between the measurement configurations for synthetic data inversion (32 transmitters and receivers) and those for experimental data inversion (8 transmitters and 241 receivers), the Green's functions derived from the experimental data differed from those obtained from synthetic data. According to the principles of multi-frequency NeuralBIM, alterations in the measurement setup necessitate adjustments in the Green's functions and data tensor dimensions, thereby requiring retraining of the multi-frequency NeuralBIM model. 

Owing to the redundancy of the 241 receivers in the experimental setup, they were uniformly downsampled to 128 receivers. Consequently, the size of the scattered field data was reduced to $128\times8$, which was then reconstructed to a $32\times32$ grid for the computation of the iterative update value of contrast. Similar to the process for synthetic data inversion, the training data was generated by combining two randomly distributed uniform cylinders within the DOI, with sizes and contrasts randomly selected from the ranges specified in Table \ref{table3}. The background within the DOI was assumed to be free space. Scatterers were randomly generated, and their EM scattering data at 3 GHz, 4 GHz, and 5 GHz were computed using MoM based on the measurement settings for experimental data inversion. Each scatterer and its corresponding EM scattering data at the three frequencies constituted a data sample, with a total of 8000 samples forming the training dataset. 

\begin{table}[htbp]
  \caption{Scatterer Parameters of Experimental Data Inversion\label{table3}}
  \centering
  \begin{tabular}{|c|c|c|c|}
  \hline
  Cylinder & Radius /$m$ & Contrast(Real) & Contrast(Imag)\\
  \hline
  $A$ & $\left[0.030,0.045\right]$ & $\left[0.2,0.8\right]$ & $0$\\ 
  \hline
  $B$ & $\left[0.010,0.025\right]$ & $\left[1.6,2.4\right]$ & $0$\\ 
  \hline
  \end{tabular}
\end{table}

\begin{figure}[htbp]
  \centering
  \includegraphics[width=0.95\linewidth]{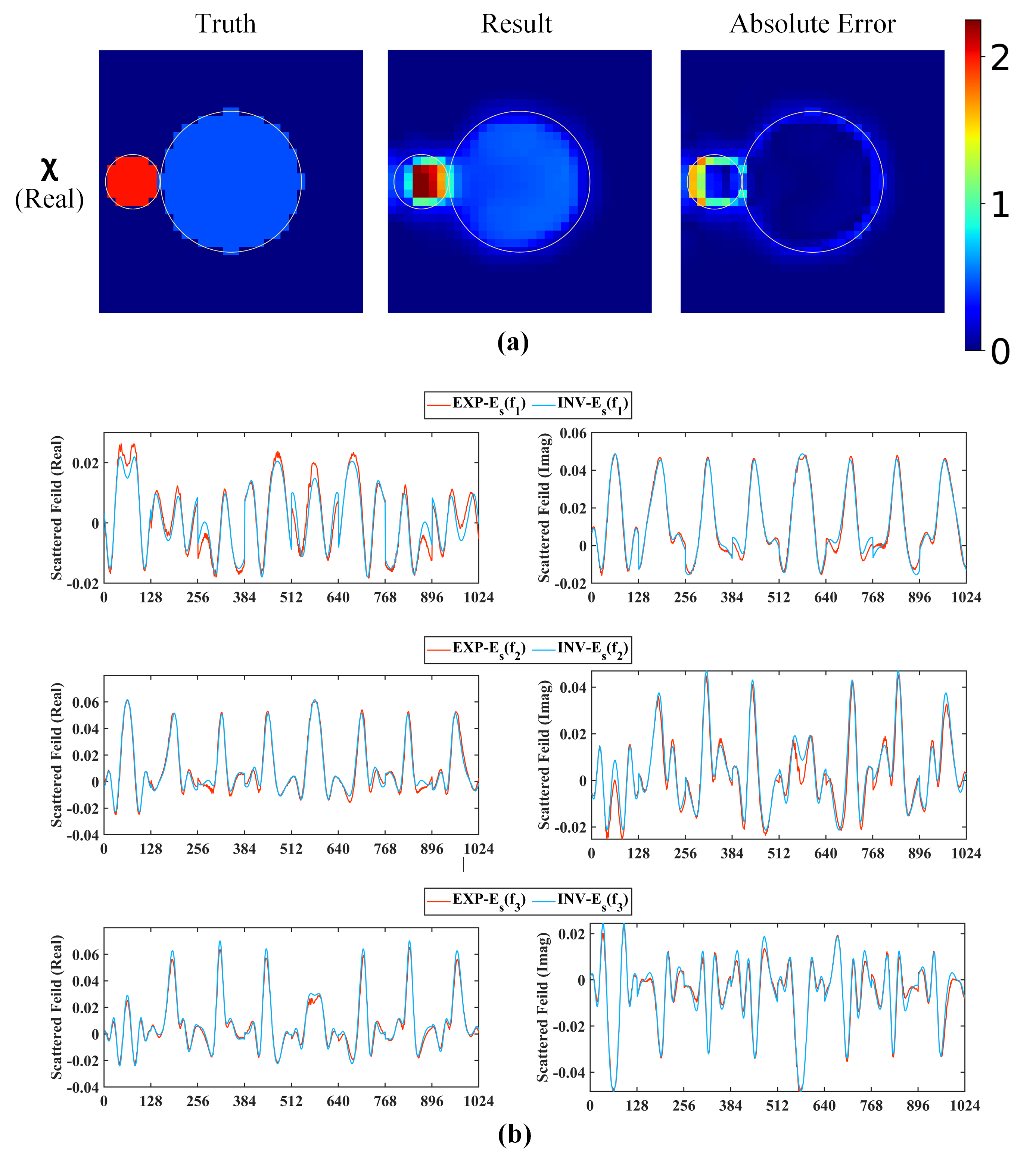}
  \caption{Evaluation of \textit{FoamDielExt} model inversion results. (a) Comparison between the ground truth and the multi-frequency neuralBIM inversion results for contrast of \textit{FoamDielExt} model. (b) Comparison of model-inversion scattered fields with experimental data scattered fields at various frequencies. $f_1$, $f_2$ and $f_3$ represent 3GHz, 4GHz, and 5GHz, respectively. }
  \label{fig12}
\end{figure}

\begin{figure}[htbp]
  \centering
  \includegraphics[width=0.95\linewidth]{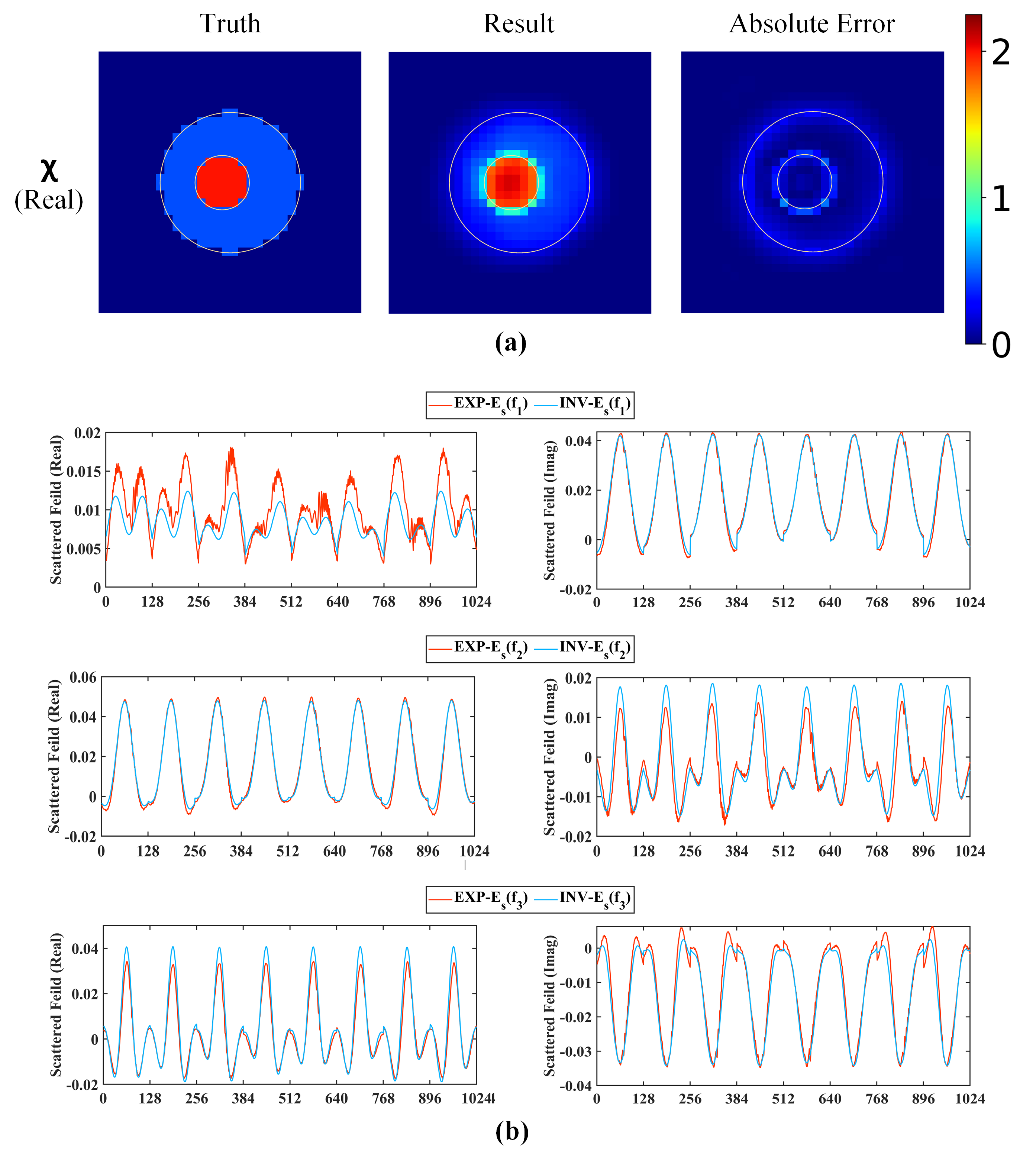}
  \caption{ Evaluation of \textit{FoamDielInt} model inversion results. (a) Comparison between the ground truth and the multi-frequency neuralBIM inversion results for contrast of \textit{FoamDielInt} model. (b) Comparison of model-inversion scattered fields with experimental data scattered fields at various frequencies. $f_1$, $f_2$ and $f_3$ represent 3GHz, 4GHz, and 5GHz, respectively. }
  \label{fig13}
\end{figure}

Upon training, the selected experimental measurement data was utilized to invert the \textit{FoamDielInt} and \textit{FoamDielExt} models. \hyperref[fig12]{Fig. 12(a)} illustrates the comparison between the inversion results of the \textit{FoamDielExt} model using multi-frequecy NeuralBIM and the actual values, demonstrating close agreement in shape and contrast. \hyperref[fig12]{Fig. 12(b)} presents the comparison between the scattered field data inverted by the model and the experimental measurement data, indicating a high degree of similarity. \hyperref[fig13]{Fig. 13(a)} shows the comparison between the inversion results of the \textit{FoamDielInt} model using multi-frequency NeuralBIM and the true values, again indicating a good inversion effect. \hyperref[fig13]{Fig. 13(b)} reveals that the scattered field data inverted by the model closely matches the experimental measurement data. The successful inversion of experimental data further corroborates the outstanding performance of the multi-frequency NeuralBIM.

\section{Conclusion}\label{section6}
In this work, we propose a deep learning-based method for solving the multi-frequency EM ISP. By combining deep learning techniques, particularly ResNet, with EM physics principles, we have successfully developed a multi-frequency NeuralBIM. The proposed multi-frequency NeuralBIM integrates multi-task learning techniques with the efficient iterative inversion processes of NeuralBIM to construct a comprehensive multi-frequency iterative inversion model. This model employs a multi-task learning approach, guided by homoscedastic uncertainty, to adaptively allocate the weights of each frequency's data during training. Additionally, an unsupervised learning method guided by physics principles is used to train the multi-frequency NeuralBIM model. This method is constrained by the physical laws of the ISP and can be used for training without requiring contrast and total field labeled data.
The effectiveness of the multi-frequency NeuralBIM is demonstrated through the validation of both synthetic and experimental data. The results show that this method not only enhances the accuracy and computational efficiency of solving ISPs but also exhibits strong generalization capabilities and noise immunity.

The multi-frequency NeuralBIM proposed in this article offers an effective solution for EM ISPs and explores a novel inversion technique for multi-frequency EM data. The successful implementation of this method introduces a new perspective and tool for solving EM ISPs, highlighting the potential and practical applications of combining deep learning techniques with physical laws in EM analysis.

\bibliographystyle{IEEEtran}
\bibliography{IEEEabrv,CITES}

\vfill

\end{document}